\documentclass[twocolumn,showpacs,pre,aps,superscriptaddress]{revtex4}
\usepackage[utf8]{inputenc}
\usepackage[english]{babel} 
\usepackage{graphicx} 
\usepackage{listings} 
\usepackage{amssymb} 
\usepackage{amsmath} 
\usepackage{longtable}
\usepackage{multirow} 
\usepackage{booktabs}
\usepackage{pst-all} 
\usepackage{pstricks-add}
\usepackage{units} 
\usepackage{listings} 
\usepackage{color}
\usepackage{subfigure}

\begin{document} 

\title{Transport in topologically disordered one-particle, tight-binding models}

\author{Abdellah Khodja}

\email{akhodja@uos.de}

\affiliation{Fachbereich Physik, Universit\"at Osnabr\"uck,
             Barbarastrasse 7, D-49069 Osnabr\"uck, Germany}

\author{Hendrik Niemeyer}

\email{hniemeye@uos.de}

\affiliation{Fachbereich Physik, Universit\"at Osnabr\"uck,
             Barbarastrasse 7, D-49069 Osnabr\"uck, Germany}

\author{Jochen Gemmer}

\email{jgemmer@uos.de}

\affiliation{Fachbereich Physik, Universit\"at Osnabr\"uck,
             Barbarastrasse 7, D-49069 Osnabr\"uck, Germany}

\begin{abstract}
 We aim at quantitatively  determining transport parameters like conductivity, mean
free path, etc., for simple models of spatially completely disordered quantum systems, comparable to the systems which are sometimes referred to as Lifshitz models. While some low-energy eigenstates in such models always show Anderson localization, we focus on models for which most states of the full spectrum are delocalized, i.e.,  on the metallic regime. For the latter we determine transport parameters in the limit of
high temperatures and low fillings using linear response theory. The
Einstein relation (proportionality of conductivity and diffusion coefficient) is
addressed numerically and analytically and found to hold. Furthermore, we find the transport behavior for some models to be in accord with a Boltzmann equation, i.e., rather long mean free paths, exponentially decaying currents, while this does not apply to other models even though they are also almost completely delocalized.
\end{abstract}


\maketitle

\section{introduction}
A large part of electronic transport theory on disordered systems is based on
spatially ordered, periodic crystal structures to which disordered impurities,
distortions, on-site potentials, etc., are added. As long as  the effect of these
disordered addends is weak, transport analysis may be performed by mapping the
electronic dynamics onto a Boltzmann equation in which the Bloch states of the
periodic part of the model correspond to free particles. This concept has been
rigorously derived in Ref. \cite{Erdos} and quantitatively applied in, e.g.,  Refs.
\cite{Mertig1987, Dekker1998, Papanikolaou1994, Vojta1992}. Whenever the influence
of the disordered part becomes large the execution of this approach becomes
challenging \cite{Mertig1987}. Quantitative results on transport in strongly
disordered 3D, one-particle quantum systems appear to be rare, some results on the
Anderson model have been reported in Refs. \cite{Steinigeweg2010, Markos2006, Kramer1993}.
In the paper at hand transport theory is approached from the opposite side: We
consider models which do not feature any spatially ordered structure whatsoever, and this 
model class is sometimes referred to as ``Lifshitz models.'' We
do not focus on weak random potentials but start from tight-binding models, the
sites of which are spatially distributed completely at random. This is incorporated
into the quantum models by means of distance-dependent hopping terms in the
tight-binding model. We find that reliable results on transport
properties of such extended, 3D, disordered models in the high-temperature limit may be obtained using standard
linear response theory and numerically exact diagonalization of finite samples
comprising about 17000 sites.

 The investigation at hand  addresses transport in substantially  disordered systems but within the delocalized 
energy regime (i.e., no thermally activated hopping transport within the localized regime), thus, the models  cannot  be 
viewed as models for, say, transport in real amorphous silicon \cite{Shklovskii1984}. The discussed type of transport may occur in strongly 
doped but weakly compensated semiconductors or amorphous metals. However, rather than modeling realistic 
systems in great detail, we focus on more general features of the transport dynamics. While the accepted picture appears to be that 
transport phenomena within the delocalized regime in disordered systems may generally be described using a Drude or 
Boltzmann approach \cite{Altshuler1985}, we find 
that this is not necessarily the case. Close to the Anderson transition there appears to be a regime in which the electrons 
are already delocalized but their transpoprt dynamics seems to be incompatible with a Boltzmann equation.

The paper at hand is organized as follows: After introducing our models in Sec.
\ref{sec-models} we identify delocalized regimes in those in Sec. \ref{loc}. This
analysis is not meant to be an exhaustive and detailed investigation of localization
in Lifshitz models, it only serves to identify the regime in which
quantitative transport investigations may reasonably be performed. In Sec.
\ref{secconduct} the conductivity at high temperatures and low fillings for a
variety of models is numerically computed on the basis of linear response theory.
Diffusion coefficient and Einstein relation are addressed, both analytically and
numerically, in Sec. \ref{seceinstein}.
Some features of the above findings on transport behavior indicate that the models
exhibit two different types of transport behavior even in the delocalized regimes:
``Boltzmann transport'' and `` non-Boltzmann transport.'' This finding is worked out in some detail in Sec. \ref{transtype}
through consideration of a mean free path. We close with a summary and conclusion in Sec. 
\ref{sumcon}.

\section{Topologically disordered tight-binding models}
\label{sec-models}
The models we investigate are three-dimensional, one-particle tight-binding models. The
sites at which the particle may be found, however, are not located periodically in
space; rather, are they distributed completely at random. Such models have been thoroughly
investigated for spectral properties, etc.,  by Lifshitz  \cite{Lifshitz1964}. However, while 
the focus there is on the density of states and the transition from the localized to the metallic 
regime, we focus on transport within the metallic regime. The finite samples of these
models on which our investigations are based are generated as follows: A cube of
volume $L^3=N$ in real space is defined. Then a set of $N$ position vectors
$\vec{x}_j$ are drawn at random by drawing each coordinate of each vector
independently from a uniform distribution on the interval $[0,L]$. This guarantees a
uniform site distribution with unit density. Now a tight-binding model with hopping or orbital overlap
terms is defined as 
\begin{equation}
\label{ham}
\hat{H}=\sum_{j k} H_{jk} \hat{a}^{\dagger}_j\hat{a}_k  
\end{equation}
where $ \hat{a}_i^{\dagger},\hat{a}_i$ denotes the annihilation and creation operators.
The function $H_{jk}$ describes the dependence of the overlap terms on the positions
of the respective sites.  We  consider isotropic overlap, thus, $H_{jk}$ essentially
depends on the distance between site $j$ and site $k$. Generally we assume $H_{jk}$
to be decreasing with increasing site distances; however, the overlap terms  will not be
limited to nearest neighbors. Since we impose periodic boundary conditions
(eventually in order to keep finite-size effects as small as possible) the distance
$s_{jk}$ is a somewhat complex function. It may be defined as 
\begin{equation}
s_{jk}:=\sqrt{d^2_{jk}(x)+d^2_{jk}(y)+d^2_{jk}(z)} 
\end{equation}
where the $d$'s are essentially the Cartesian components of $(\vec{x}_j- \vec{x}_k)$
. Due to periodic boundary conditions they are now defined as  
\begin{equation} 
d_{jk}(\alpha)=\begin{cases}
 |\alpha_j-\alpha_k|,  & |\alpha_j-\alpha_k|<\frac{L}{2}\\
 L-|\alpha_j-\alpha_k|,  &|\alpha_j-\alpha_k|>\frac{L}{2}
\end{cases}
\end{equation}
where $\alpha$ is one of the Cartesian coordinates, i.e., $\alpha=x,y,z$. Thus, the
distance $s_{jk}$ is essentially the shortest distance between the sites $j,k$ under
periodic closure of the sample. With this definition of the distance we  now specify
three model\_types as follows:

{\it{Model type I}}: The overlap terms as entering the Hamiltonian in (\ref{ham}) are
taken to decrease exponentially with the distance, i.e., 
\begin{equation}
 H^{I}_{jk}:= \exp\left(\frac{-3 s_{jk}}{\tilde{l}}\right)
\end{equation}
where $\tilde{l}$ is a parameter that equals  the mean overlap length, i.e., this and
all following overlap terms are  constructed such that
\begin{equation}
\label{meanhop}
 \frac{1}{N}\sum_{jk} s_{jk} |H^{I,II,III}_{jk}|=\tilde{l}
\end{equation}
This specific model type has been chosen since its Anderson transition with respect
to $\tilde{l}$ has been discussed in the literature and, thus, the corresponding value
$\tilde{l}\approx 0.6$ is fairly  well known, cf. Ref. \cite{Bauer1988} and references
therein. (Note that the definition of the parameter that controls the overlap length
in Ref. \cite{Bauer1988} differs  slightly  from the one at hand.) However, as will
become clear below, it is particularly difficult to obtain quantitative results on
transport behavior for this specific model\_type. Thus, we introduce another
model\_type for which results are much less affected by finite-size effects.

{\it{Model type II}}: Here we define the overlap terms as decreasing as a Gaussian
with the distance, i.e., 
\begin{equation}
 H^{II}_{jk}:= \exp\left(\frac{-4 s_{jk}^2}{\pi\tilde{l}^2}\right)
\end{equation}
again the function is constructed in such a way that (\ref{meanhop}) holds. This
model is interesting since it shows, as will become clear below, a transition from 
transport behavior comparable to Brownian motion (non-Boltzmann transport) to transport
dynamics as occurring in a crystal with some impurities (Boltzmann transport) with
increasing $\tilde{l}$. It turns out that this transition only occurs if the phases
of the overlap terms are non random.  In order to demonstrate this we
investigate a third type.

{\it{Model type  III}}: Here we also define the absolute values of the overlap terms as
decreasing as a Gaussian with the distance; however, we allow for random phases
\begin{equation}
H^{III}_{jk}:= \exp\left(\frac{-4 s_{jk}^2 }{\pi\tilde{l}^2}  +i\phi_{jk}\right)
\end{equation}
where the $\phi_{jk}$, for say, $j>k$ are real random numbers drawn independently
from the interval $[0,2\pi]$. Of course, to guarantee hermiticity of the Hamiltonian
the  $\phi_{kj}$ have then to be chosen as $\phi_{kj}=-\phi_{jk}$. This leaves the
absolute values unchanged such that (\ref{meanhop}) still holds. In some sense this
model is even more random than model\_type I and model\_type II and, as it turns out,
shows non-Boltzmann transport  only.

All above models but especially model-type III bare a similarity with the ``banded
random matrix models'' as discussed in, e.g., Ref. \cite{Erdoes2011}, in which the
overlap terms that connect sites below a certain distance are simply
chosen at random, all other overlap terms  are set to zero. In  Ref. \cite{Erdoes2011} it
is reported that those models exhibit diffusive behavior in 3D on a relevant time
scale if the characteristic overlap length is sufficiently long. Although the models
at hand differ somewhat, our results are in principle in accord with the
findings in Ref. \cite{Erdoes2011}.

The above models are all isotropic, i.e., the hopping amplitudes depend on the
distance of the respective sites only. This, however is no crucial prerequisite for
all the below investigations. Anisotropic disordered systems like, e.g., discussed in
Ref. \cite{Rodriguez2012} may be analyzed in the same way.

\section{Detection of delocalized regimes as a basis for transport investigations}
\label{loc}
The main focus of the paper at hand is the quantitative analysis of macroscopic
transport properties of the above defined models (cf. Sec. \ref{sec-models}).
However, as we are dealing with disordered models Anderson localization may occur
which affects the transport properties severely. Localized states are basically energy
eigenstates that feature  spatial probability distributions with finite spreads,
even if the model is infinitely large. For a given 3D model at a given energy the
eigenstates are either localized or extended. Usually the delocalized states are in
the center of the spectrum (or the bands) while localization occurs at the edges
\cite{Kramer1993, Abrahams2010, Knief1998}. If some model parameter, e.g., the
magnitude of disordered on-site potentials, hopping lengths, etc., is tuned in some
direction the localized part of the spectrum may increase such that at some point
all states are localized (Anderson transition). If the parameter is tuned to 
the opposite extreme it is expected that (in 3D) in the limit practically all
states at all energies become delocalized  \cite{Kramer1993}. At an energy regime at
which states are localized all macroscopic transport coefficients such as diffusion
constant, conductivity, etc., vanish, i.e., the system behaves as an isolator on the
large scale (this refers to the isolated system, i.e., no phonon-assisted, thermally activated transport like discussed, e.g., in Ref. \cite{Shklovskii1984} is considered here). In the work at hand we are interested in the transport behavior of the delocalized regime only. In order to make sure that the changes in transport behavior with model parameters which we investigate below are not simply due to the onset of localization we first aim at finding models and parameter regimes which are almost completely delocalized. To this end we have to consider the spectrum and the delocalized part of it. The precise
determination of the mobility edge (precise energy which seperates localized from delocalized regimes) is a formidable task of its own. In the context
of the Anderson model it has been approached by sophisticated techniques such as
transfer matrix methods, Thouless scaling, fractal analysis, etc.\cite{Abrahams2010, Brndiar2006}. For our purposes
it suffices to have less precise information on the mobility edge since we only
intend to find out whether the biggest part of the spectrum is delocalized.
Thus, we employ a rather simple criterion to determine the approximate position of
the mobility edge. This criterion is based on the inverse participation number
\begin{equation}
I(E):=\sum_{i=1}^{N}|\psi(E,i)|^{4}  
\end{equation}
Here $\psi(E,i)$ is the amplitude at site $i$ of an energy eigenstate with energy
$E$. Localization analysis based on the inverse participation number is well established. It has been used even for rather detailed investigations in the Anderson model \cite{Brndiar2006}, as well as for topologically disordered systems
\cite{Atta-Fyn2004}
Here we suggest a very simple ``delocalization indicating criterion'': we guess that states
at an energy $E$ are surely delocalized if  $I(E)< 90/N$ for a finite but large enough
model of size $N$. This guess appears to be in reasonable accord with various more carefully derived results on the mobility edge from the literature. It has furthermore the advantage of being numerically cheap since it turns out that averaging over about five random models described by the same model parameters suffices to identify  $E_{\text{deloc.}}$ from $I(E_{\text{deloc.}})=90/N$ with reasonable accuracy. We find (as expected) that this equation has two solutions, thus the regime between those two energies is delocalized.

\begin{figure}[htbp]
    \subfigure[]{\includegraphics[width=7.5cm]{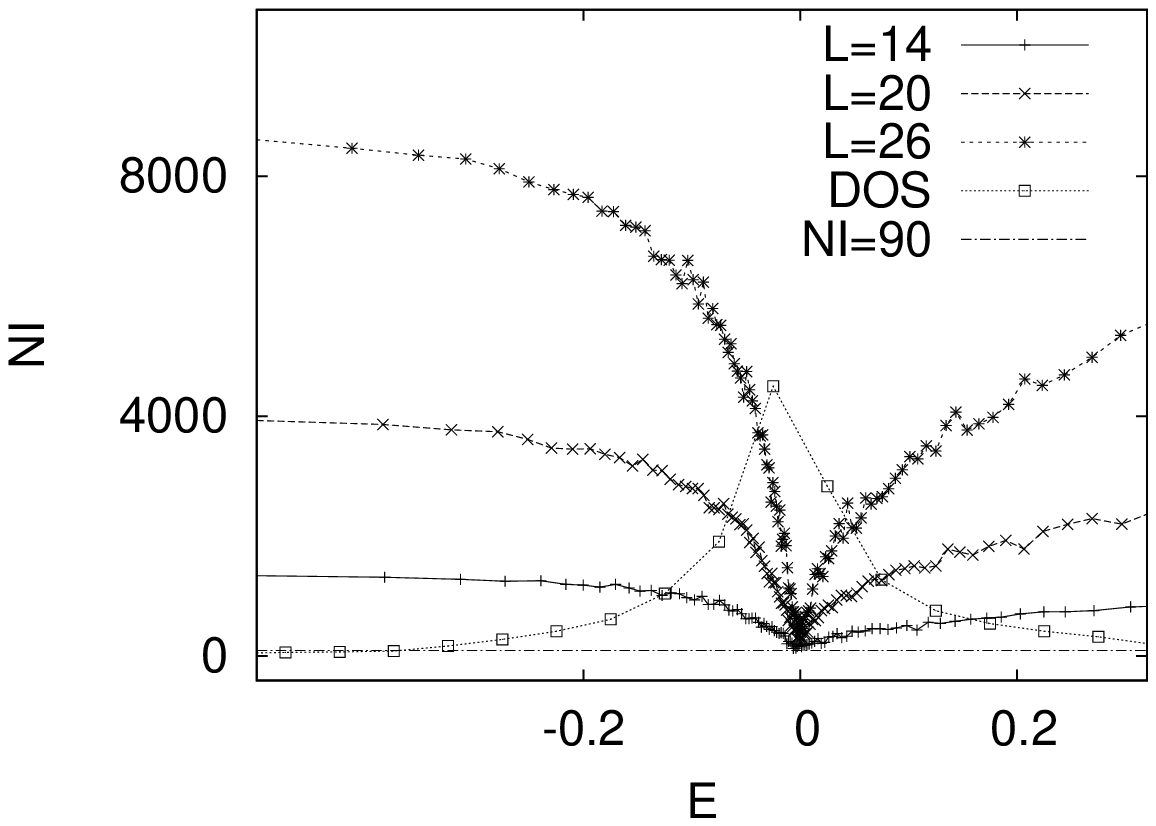}
\label{fig-1a}
}
    \subfigure[]{\includegraphics[width=7.5cm]{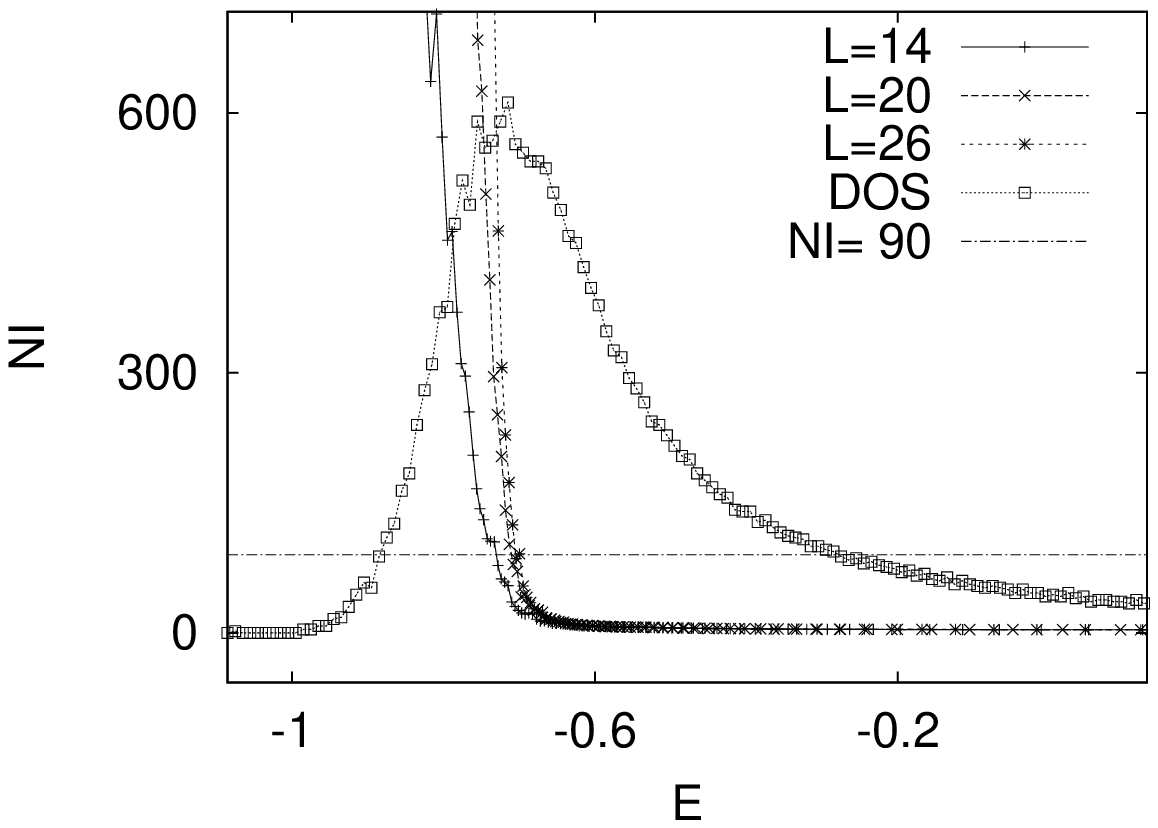}
\label{fig-1b}
}
\caption{Scaled inverse participation number $N I(E)$ and density of states (DOS)
for model\_type I. (a) Corresponds to mean overlap length $\tilde{l}$=0.6 which
appears to be at the Anderson transition; (b) corresponds to mean overlap length
$\tilde{l}$=4; an extended delocalized regime appears to exist.}
\label{fig-1}
\end{figure}

\begin{figure}[htbp]
\subfigure[]{\includegraphics[width=7.5cm]{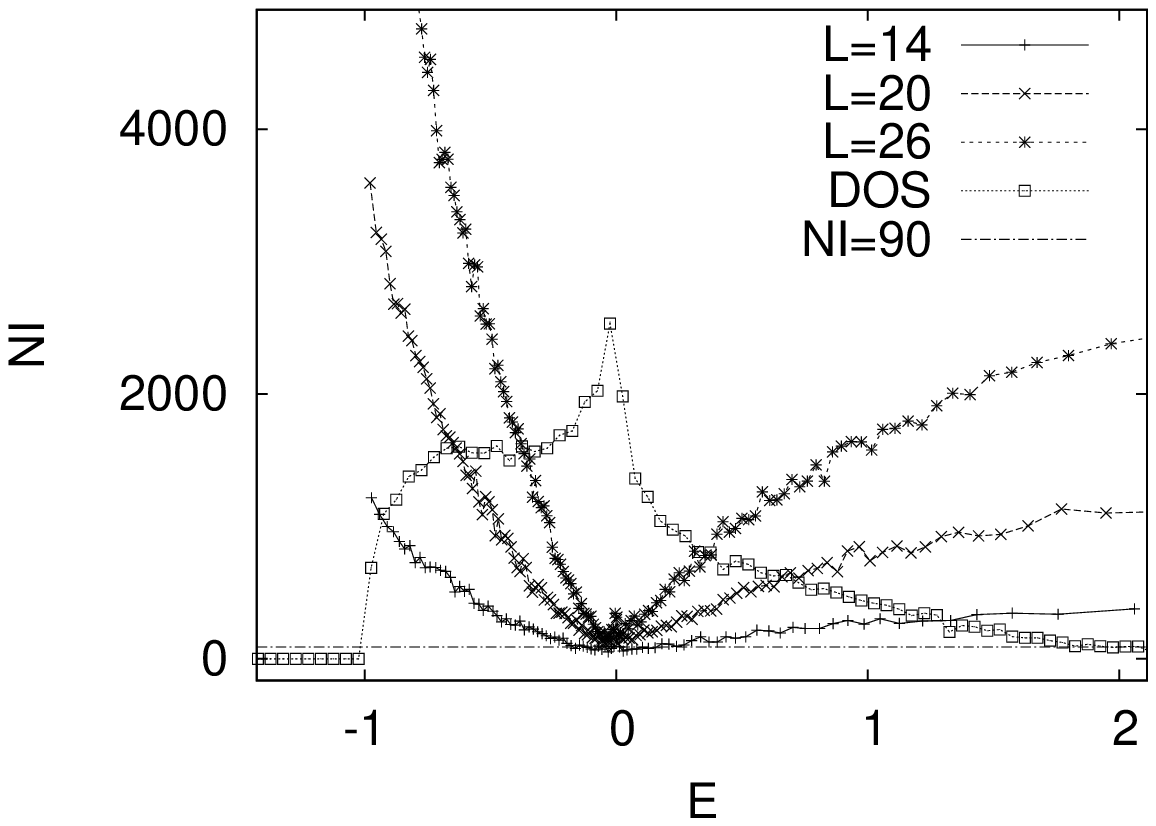}
\label{fig-2a}
}
\subfigure[]{\includegraphics[width=7.5cm]{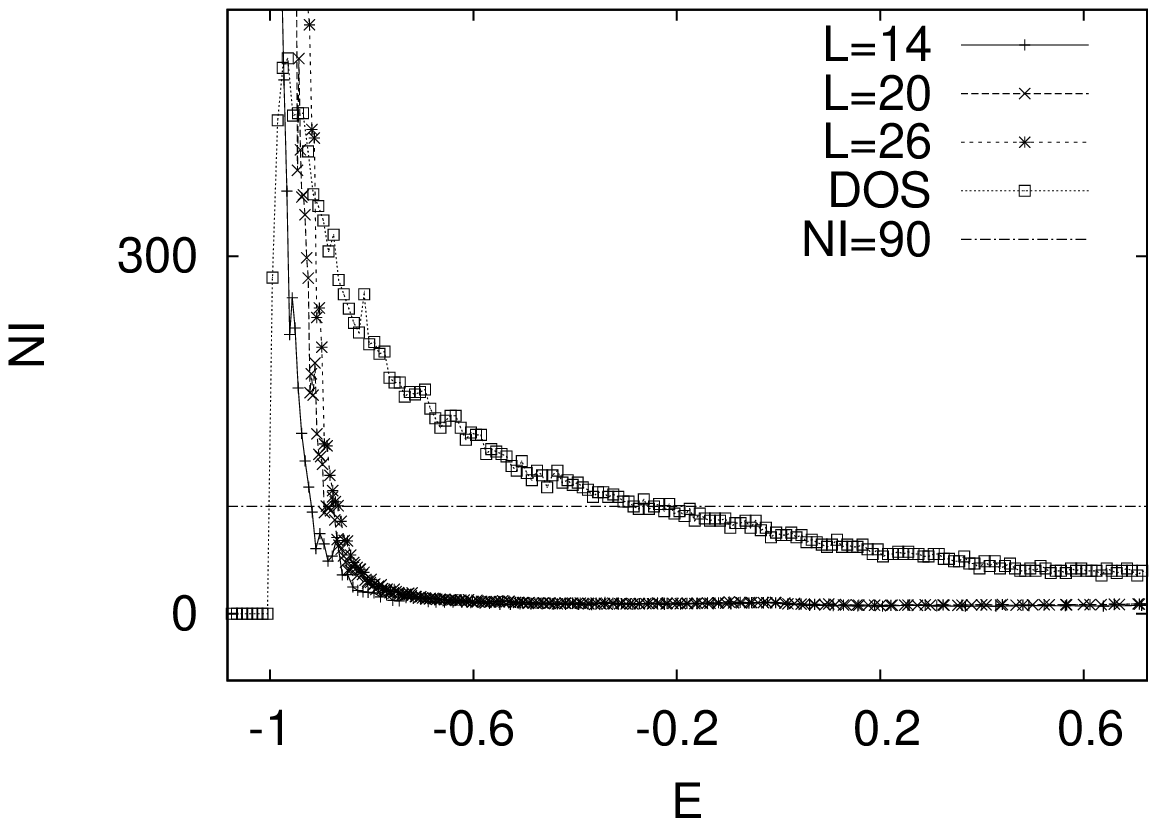}
\label{fig-2b}
}
\caption{Scaled inverse participation number $N I(E)$ and density of states (DOS)
for model\_type II. (a) Corresponds to mean overlap length $\tilde{l}$=0.7 which
appears to be at the Anderson transition; (b) corresponds to mean overlap length
$\tilde{l}$=1; an extended delocalized regime appears to exist. }
\label{fig-2}
\end{figure}
\begin{figure}[htbp]
\subfigure[]{\includegraphics[width=7.5cm]{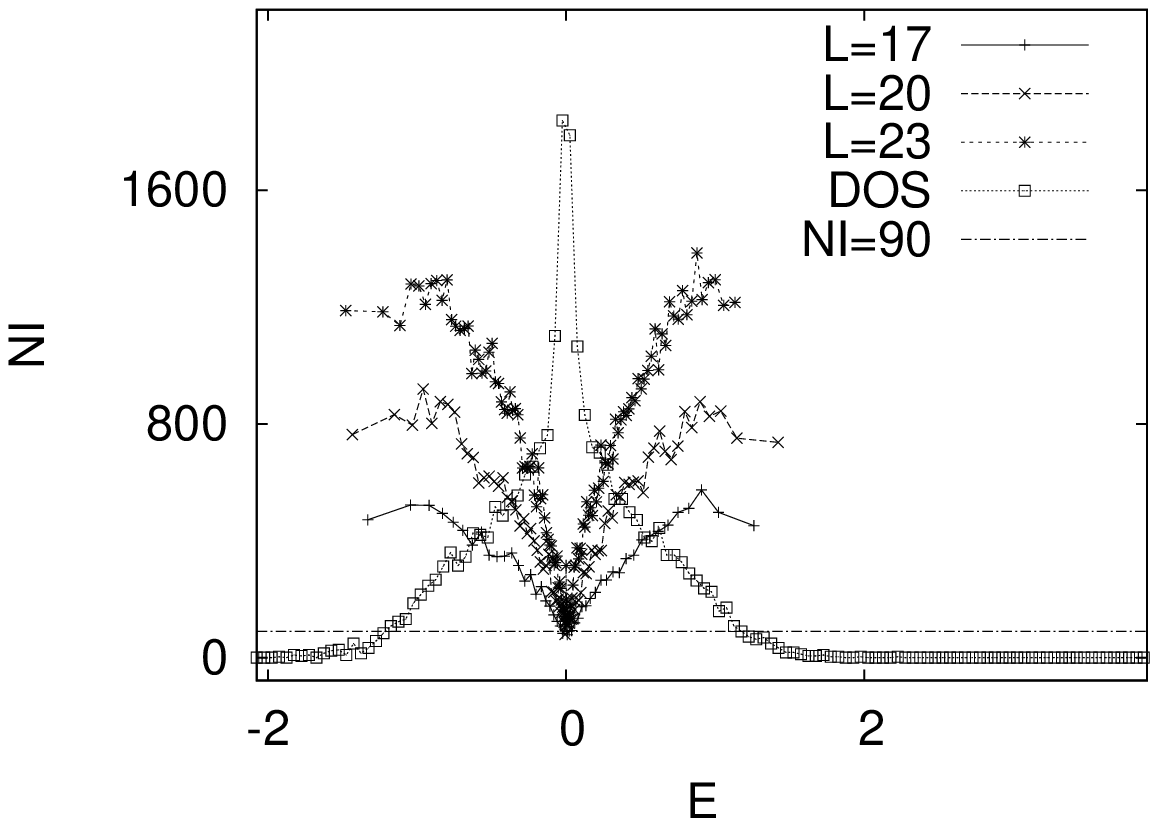}
\label{fig-3a}
}
\subfigure[]{\includegraphics[width=7.5cm]{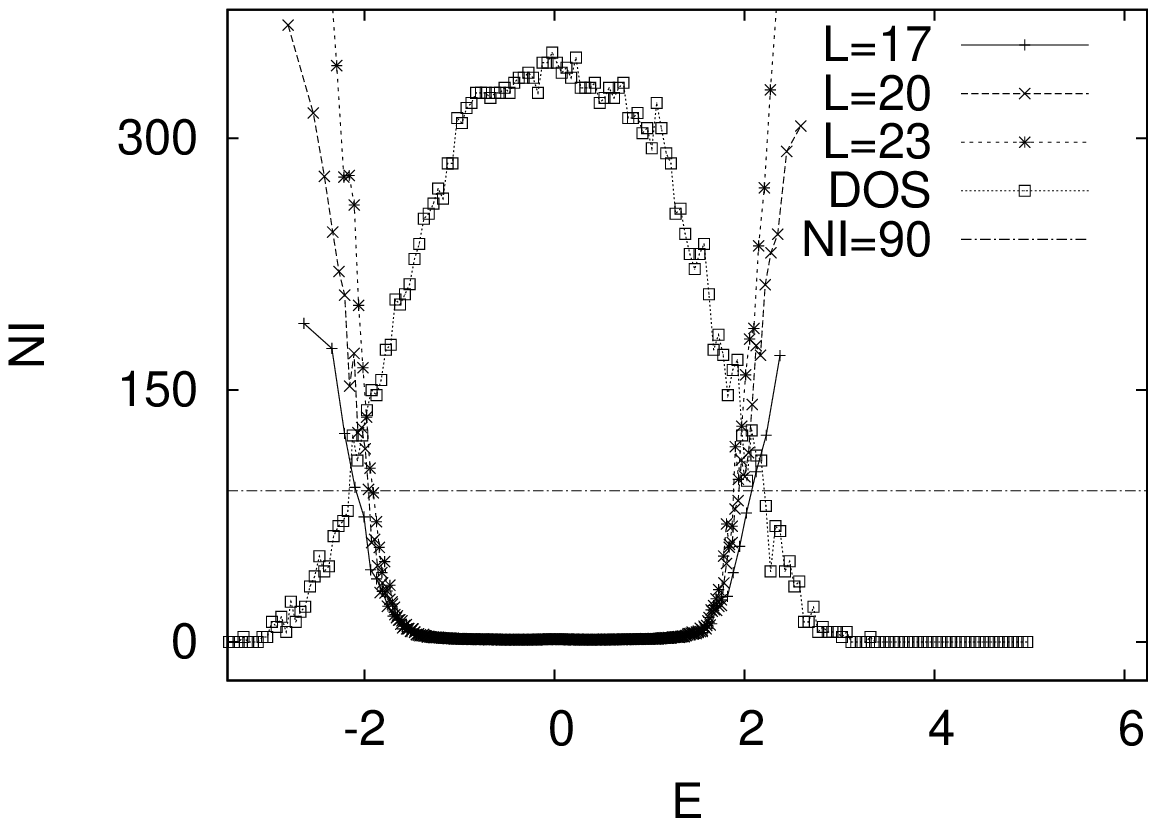}
\label{fig-3b}
}
\caption{Scaled inverse participation number $N I(E)$ and density of states (DOS)
for model\_type III. (a) Corresponds to mean overlap length $\tilde{l}$=0.6  which
appears to be at the Anderson transition; (b) corresponds to mean overlap length
$\tilde{l}$=1;  an extended delocalized regime appears to exist. }
\label{fig-3}
\end{figure}
We illustrate our ``delocalization criterion'' in  Figs. \ref{fig-1},\ref{fig-2}, and \ref{fig-3}, which refer to the model\_types I, II and III, respectively. The Figs.
show a ``scaled inverse participation number,'' i.e., $NI(E)$ as a function of $E$
for various $N$ together with the density of states (DOS), given in arbitrary units.
(If not mentioned otherwise all data presented in this paper are averaged over five
random implementations of the addressed model as defined in Sec. \ref{sec-models}.
However, this averaging appears not to be crucial, typically the data for various
implementations of the same model looks very similar. Wherever the data are presented
as a histogram of course a further averaging over an appropriate bin size is
performed.)

\begin{figure}[htbp]
\centering
\includegraphics[width=7.5cm]{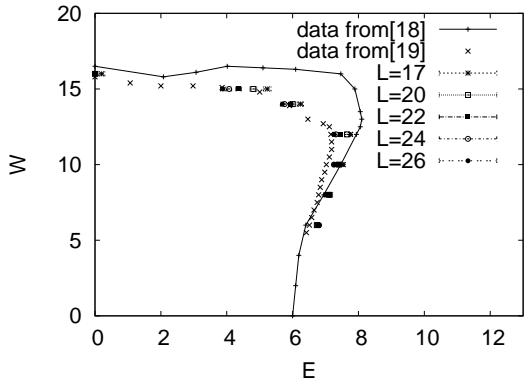}
\caption{Results on the mobility edge in the 3D Anderson model; $W$ quantifies the
degree of disorder. Compared are results from the ``$90/N''$\_criterion for different
sample sizes to results from more refined methods from the literature
\cite{Grussbach1995, Bulka1987}. The results from our $90/N$\_criterion appear to
converge reasonably against the results from Ref. \cite{Bulka1987}. Error bars for the
$90/N$\_criterion indicate variations arising from different random implementations
of the Anderson model featuring the same degree of disorder.}
\label{fig-4}
\end{figure}

Figures.  \ref{fig-1a}, \ref{fig-2a}, and \ref{fig-3a} correspond to mean overlap lengths
 $\tilde{l}$ which are only very slightly above the Anderson transition. Accordingly, the regime in which the  $NI(E)$ coincide is very narrow and appears to yield $NI(E)\approx90$. The value for Fig. \ref{fig-1a} is in very good agreement with results on the Anderson transition for the same model reported in \cite{Bauer1988} and references therein. Figures. \ref{fig-1b}, \ref{fig-2b}, and \ref{fig-3b} correspond to mean overlap lengths $\tilde{l}$ substantially above the Anderson transition. Correspondingly, there are extended regimes to which $NI(E)\leq 90$ applies. Furthermore, for large-enough $N$  the scaled inverse participation numbers to which  $NI(E)\leq 90$ 
applies appear to become independent of  $N$ which indicates that states within this regime are indeed delocalized. For some models and some energies we additionally computed the mobility edge from the more costly method described in Ref. \cite{Brndiar2006}. It turns out that the deviations between the so-computed mobility edge and our $E_{\text{deloc.}}$  are on the order of $5\%$.   As final demostration of the validity of our delocalization criterion, we compare the energies at which  $NI(E)=90$ for the 3D Anderson models with results on the mobility edge from the  literature \cite{Grussbach1995, Bulka1987}. (Note that this is only to
``calibrate'' our criterion, i.e., determine the factor ``90.''  We will not analyze the Anderson model quantitatively
for transport; such an investigation may be found in Ref. \cite{Steinigeweg2009b}.)
The results are shown in Fig. \ref{fig-4}. Obviously there is reasonable agreement
between result from our criterion and data from the literature. The agreement appears to become better in the regime we are interested in, i.e.,  disorders $W$ where most of the eigenstates are delocalized. 

Our method is not appropriate to determine the Anderson transition or the mobility edges
with great precision because $NI(E_c)$ in the critical region is not a constant but depends on the sample size (fractal dimension). It, however, appears well suited to quickly identify models for which
the vast majority of all energy eigenstates is delocalized which is the purpose of
the investigation at hand.
It may be worth noting here that frequently, e.g., in the context of the Anderson
model, the mobility edge lies in an energy regime with a relatively low density of
states. This, however, appears to hold for the present models only for type III. For
model\_types I and II the mobility edge appears to lie at or close to the maximum
density of states, cf. Figs. \ref{fig-1b}, \ref{fig-2b}, and \ref{fig-3b}.
\begin{figure}[htbp]
\subfigure[]{\includegraphics[width=7.5cm]{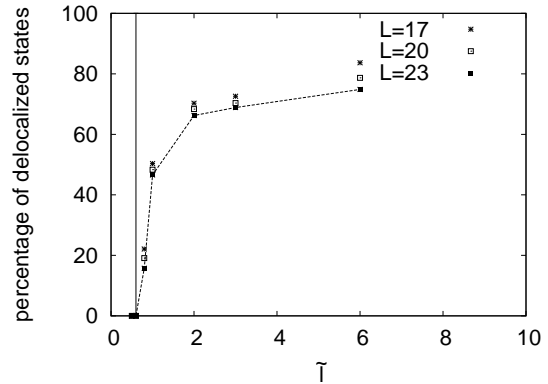}
\label{fig-5a}
}
\subfigure[]{\includegraphics[width=7.5cm]{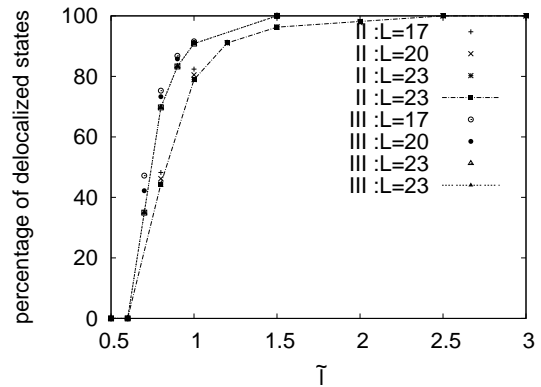}
\label{fig-5b}
}
\caption{Percentage of delocalized energy eigenstates for various mean overlap
lengths $\tilde{l}$. Panel (a) addresses  model\_type I. The vertical solid line at $
\tilde{l}=0.6$ indicates the Anderson transition as found in Ref. \cite{Bauer1988}.
Although the data appear not to be fully converged for some $\tilde{l}$ it is
obvious that about $30\%$ of the spectrum remain localized up to  $ \tilde{l}=6$.
Panel (b) addresses model\_types II and III. Although the data appear not to be fully
converged for some $\tilde{l}$ close to the Anderson transition, it is obvious that
less than $10\%$ (and decreasing) of the spectrum are localized for  $ \tilde{l} >
1.3$. }
\label{fig-5}
\end{figure}
After those preliminary considerations we now simply calculate the percentage of all
energy eigenstates for which   $NI(E)\leq 90$, according to the previous considerations  we expect this to be close to or a little smaller then the percentage of all eigenstates that are delocalized. This is
done for all three model types and various mean overlap lengths. The result is
displayed in Fig. \ref{fig-5}. Obviously, the Anderson transition for all three model
types occurs, very roughly, around $\tilde{l} \approx 0.6$. However, whereas for
model\_types II and III almost all states are delocalized for  $\tilde{l} \geq 1.3$
[Fig. \ref{fig-5b}] a substantial fraction of states remains localized up to
$\tilde{l} \approx 6$ for model\_type I [Fig. \ref{fig-5a}]. Since it is a reasonable
assumption that, using numerical diagonalization, reliable transport constants  may
only be  obtained for sample sizes that are large compared to the mean overlap
length, we do not pursue the analysis of model\_type I any further for we are
numerically limited to sample sizes of $L=26$.
\section{Conductivity at low fillings and high temperatures}
\label{secconduct}
First, we investigate the conductivity of model\_types II and III for various hopping
lengths. We employ linear response theory, i.e., the  Kubo formula. In the limit of
high temperatures and low fillings (routinely described within the framework of the
grand-canonical ensemble) the dc\_conductivity is given as:
\begin{equation}
\label{kubo}
\sigma_{dc}= \sigma(t\rightarrow \infty), \quad   \sigma(t) =
\frac{f}{T}\int_0^{t}\frac{1}{N}\text{Tr}\{ \hat{J}(t')\hat{J}(0)\} dt'   
\end{equation}
\cite{Kubo1991, Jaeckle1978}, where  $f$ is the filling factor (mean number of
particles  per site at equilibrium), $N$ denotes the number of sites in the sample
(since we work at unit spatial site density $N$ also equals the volume), trace and
current operators refer to the one-particle sector only, and, furthermore, $J(t)$ denotes
the current operator in the Heisenberg picture. $T$ is the temperature and we set
$k_B=1$, $\hbar=1$, and,  furthermore, we set the charges of the particles to unity, i.e.,
$q=1$. Now, of course, an appropriate current operator has to be defined. In the
context of periodic systems this is often done by considerations based on the
continuity equation for the particle density \cite{Zotos1999, Heidrich-Meisner2003,
Benz2005, Gemmer2006}. Here we choose a  definition of the current which is
based on the ``velocity'' in, say, $x$\_direction,  i.e.,  
\begin{equation}
\label{vel}
\hat{v}=i[\hat{H},\hat{x}]
\end{equation}
Here $\hat{x}$ is a $x$\_position operator and it is defined as 
\begin{equation}
\label{pos}
\hat{x}= \sum_{i=1}^{N}x_i\hat{n}_i, \quad  \hat{n}_i:= \hat{a}_i^{\dagger}\hat{a}_i
\end{equation}
where $x_i$ is the $x$\_coordinate of the position of site $i$. Thus, the operator
$\hat{v}$ may also be written as 
\begin{equation}
\label{velocity}
\hat{v}=  i\sum_{ij}(x_j-x_i)H_{ij}  \hat{a}_i^{\dagger}  \hat{a}_j     
\end{equation}
The interpretation of such an operator as velocity or current is somewhat in
conflict with periodic boundary conditions (which we impose for technical reasons).
A (slow) transition of probability from, say, the right edge of the sample ($x=L$)
to the left edge of the sample ($x=0$) would give rise to very high negative
velocities. But within the concept of periodic boundary conditions such a transition
should correspond to low positive velocities. Thus, in order to obtain a suitable
current operator, we modify the above velocity operator (\ref{velocity}) such that it
features the same structure for transitions arising from the periodic closure as it
already exhibits for transitions within the sample,
\begin{equation}
 \hat{J}=\sum\limits_{ij} J_{ij} \hat{a}^\dagger_i \hat{a}_j
\end{equation}
\begin{equation*}
 J_{ij}=\begin{cases}
 i[x_j-x_i]H_{ij}  &|x_j-x_i|<\frac{L}{2} \\
 \mathrm{sgn}(x_j-x_i)\left[i[L-|x_j-x_i|]H_{ij}\right]  &|x_j-x_i|>\frac{L}{2} 
\end{cases}
\end{equation*}
Equipped with this definition for the current we may now simply calculate the
current auto correlation function as appearing in  (\ref{kubo}). We do so using
standard numerically exact diagonalization routines. Within reasonable computing
time we are able to treat samples up to a size of $L=26$. It turns out that this
appears to be sufficient to render finite-size effects for a range of models
negligible. In order to be able to compare the key features of the dynamics of  the
current auto correlation functions for various model types and sizes to each other
we compute a kind of ``normalized'' current auto correlation functions,
$j^{\prime}(t):=\text{Tr}\{ \hat{J}(t)\hat{J}(0)\}/\text{Tr}\{ \hat{J}^2(0)\}$.
\begin{figure}[htbp]
  \subfigure[]
{\includegraphics[width=7.5cm]{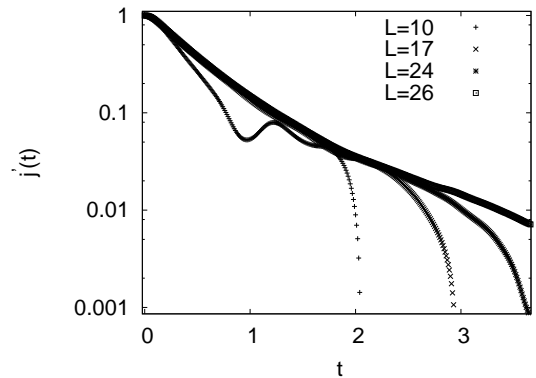}
\label{fig-6a}
}
  \subfigure[]{\includegraphics[width=7.5cm]{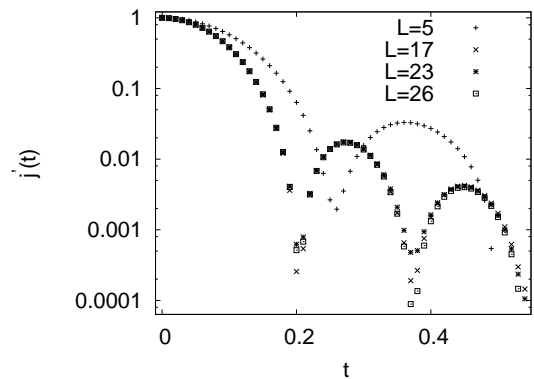}
\label{fig-6b}
}
\caption{Normalized current auto correlation function $j^{\prime}(t)$ for increasing sample
sizes $L$. Note that due to the logarithmic scale the initial dynamics is most
relevant. Panel (a) addresses model\_type II, with mean overlap length $\tilde{l}= 2.0$. The
data appear to be reasonably free of finite-size effects for $L \geq 24$. The decay
appears to be essentially exponential. Panel (b) addresses model\_type III, with mean overlap
length $\tilde{l}= 4.0$. The data appear to be reasonably free of finite-size
effects for $L \geq 17$. The decay appears to be essentially Gaussian}
\label{fig-6}
\end{figure}

\begin{figure}[htbp]
 \subfigure[]{\includegraphics[width=7.5cm]{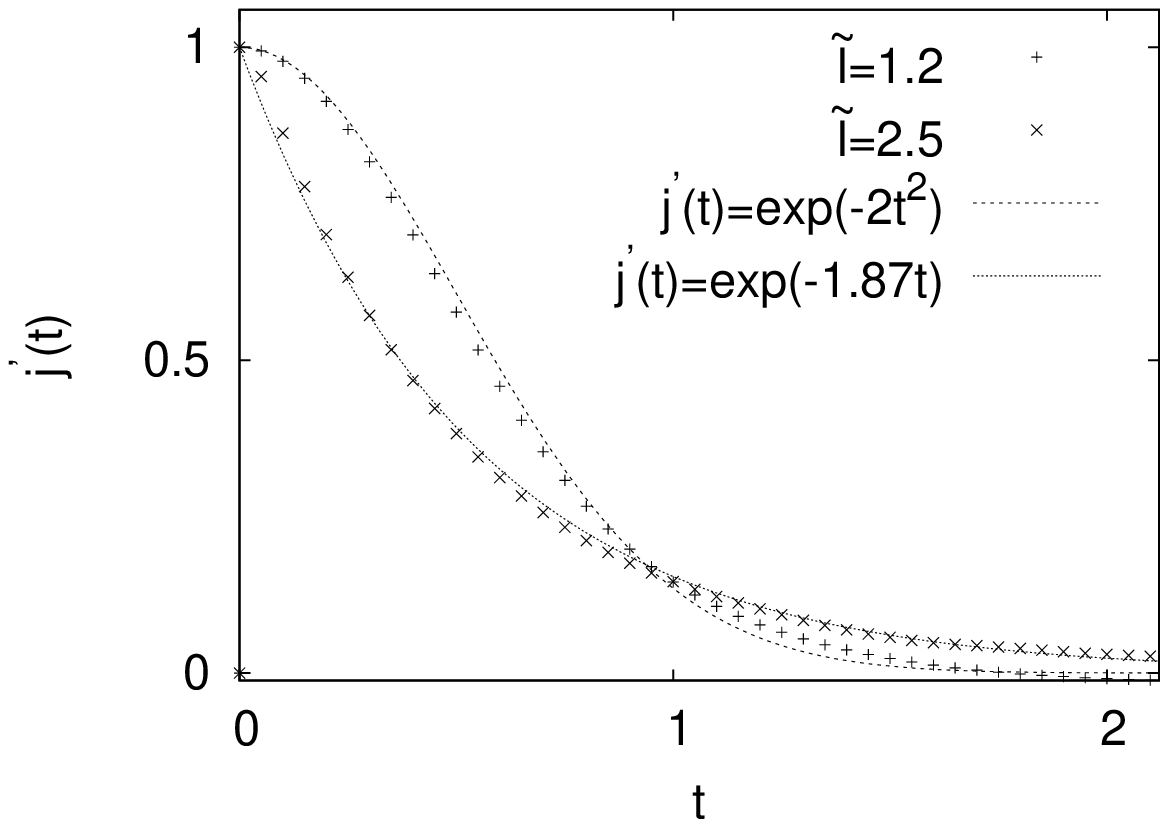}
\label{fig-7a}
}
  \subfigure[]{\includegraphics[width=7.5cm]{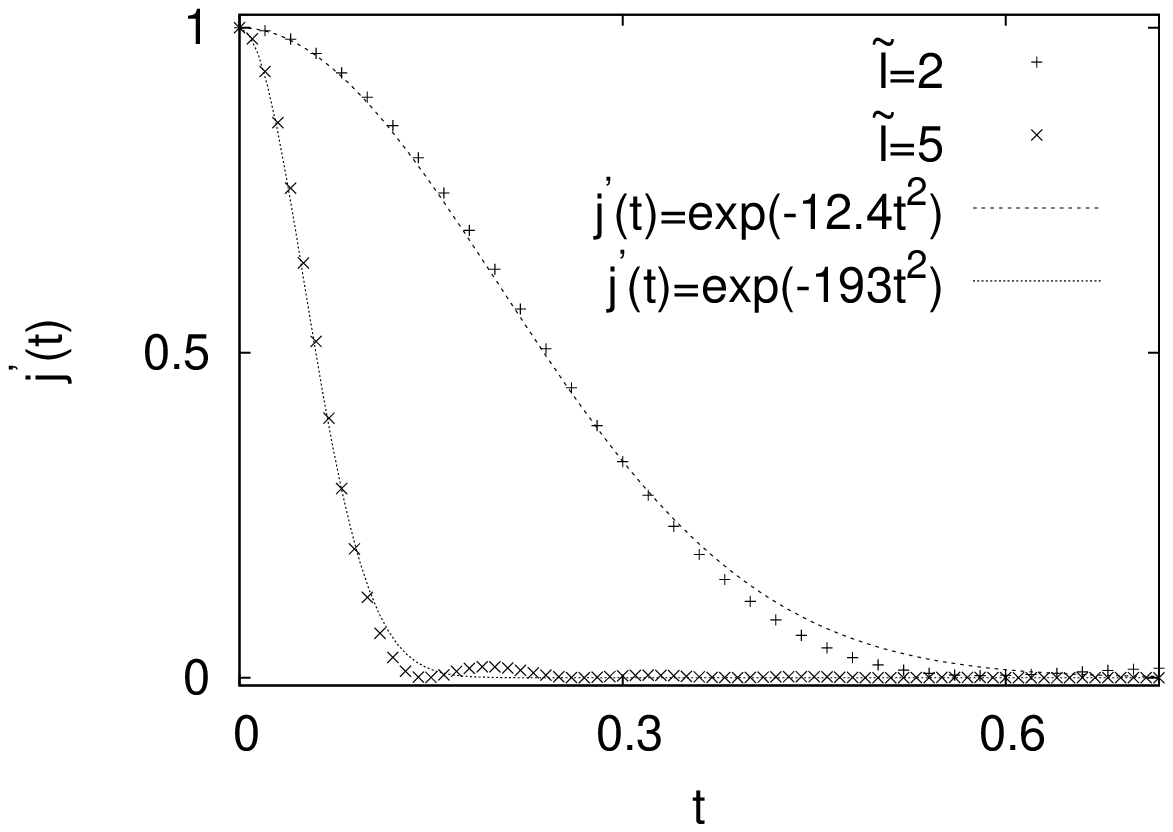}
\label{fig-7b}
}
\caption{Normalized current auto correlation function $j^{\prime}(t)$ for the longest and
shortest reasonable mean overlap lengths $\tilde{l}$ (see text), sample size $L=26$. Panel (a)
addresses model\_type II, with mean overlap length $\tilde{l}= 1.2$ and $\tilde{l}= 2.5$.
The decay appears to be dominantly Gaussian for the short and dominantly exponential
for the long mean overlap lengths. Panel (b) addresses model\_type III, with mean overlap lengths 
$\tilde{l}= 2.0$ and $\tilde{l}= 5.0$. The decay appears to be essentially Gaussian
for both mean overlap lengths. }
\label{fig-7}
\end{figure}
 Figures \ref{fig-6a} and \ref{fig-6b} show $j^{\prime}(t)$ for model\_types II and III for
different sample sizes. Obviously the $j^{\prime}(t)$ coincide for the larger sample sizes
for the relevant times, i.e., for times at which $j^{\prime}(t)$ is substantially
different from zero. From this finding we conclude that in this case finite-size
effects are indeed negligible. However, whether or not  $L=26$ is sufficient to get
rid of finite-size effects depends on the model\_type and the mean overlap length.
Typically, finite-size effects are less severe for shorter mean overlap lengths. Figure.
\ref{fig-7a} shows $j^{\prime}(t)$ for model\_type II ($L=26$) for two mean overlap lengths.
$\tilde{l}= 1.3$ is the shortest mean overlap length for which most likely the
largest part of the spectrum is still delocalized [cf. Fig. \ref{fig-5b}],  
 and $\tilde{l}= 2.2$ is the longest mean overlap length for which we obtain results 
that are reliably 
unaffected by finite size effects. Figure \ref{fig-7b}  shows $j^{\prime}(t)$ for model\_type
III for an intermediate mean overlap length.  For model\_type II the relaxation
dynamics appear to undergo a transition from a Gaussian decay to an exponential
decay as the mean overlap lengths become larger, cf. Fig. \ref{fig-7a}. For model\_type III
the relaxation dynamics are more or less Gaussian at all mean overlap lengths, cf. Fig
\ref{fig-7b}. The exponential decay of the current for model\_type II at long mean overlap lengths suggests that the respective dynamics may be interpreted as the
dynamics of almost free (lattice) particles which are only weakly scattered
\cite{Jaeckle1978}.  Within the framework 
of such and interpretation the (disordered) eigenstates of the current operator
$\hat{J}$ take the role of the Bloch states in a periodic crystal. Since such a
behavior may be described on a phenomenological level by a Boltzmann equation we
call this type of dynamics ``Boltzmann transport.'' For all cases in which the current decays non exponentially (e.g., Gaussian)
the dynamics cannot be described by a standard Boltzmann equation, this occurs although this cases correspond to the metallic (delocalized) regime as well. Thus, we call this type of dynamics  ``non-Boltzmann transport.'' We elaborate on this issue in
more detail ins Sec. \ref{transtype}.
\begin{figure}[htbp]
\centering
\includegraphics[width=7.5cm]{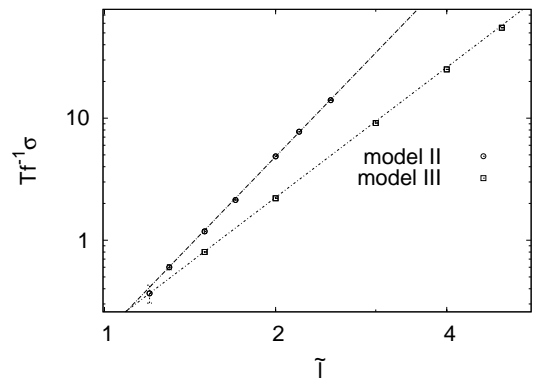}
\caption{ Scaled conductivity $T f^{-1}\sigma_{dc}$  (or diffusion constant $D$, see
text) for model\_types II and III as a function of mean overlap length  $\tilde{l}$, with
sample size $L=26$. For both model\_types the conductivity appears to scale as a
power law with $\tilde{l}$; the dashed ($---$, model\_type II) and dotted ($...$,
model\_type II) lines are the respective fits, cf. (\ref{conduct}). Whereas at
$\tilde{l}\approx 1.3$ the conductivities more or less coincide the conductivity for
model\_type II appears to increase much faster. The vertical error bars correspond to
different random implementations of the same model.}
\label{fig-8}
\end{figure}
 To conclude the considerations on conductivity we plot $\sigma_{dc} f/T$ for model\_types II and III over mean overlap length (on a double logarithmic scale); see Fig.
\ref{fig-8}. The plot clearly suggests a power-law scaling of the conductivity mean overlap length.
The corresponding fits yield, for the respective conductivities,
\begin{equation}
\label{conduct}
\sigma^{II}_{dc}=\frac{f}{T}0.17 \, \, \tilde{l}^{4.83}, \quad 
\sigma^{III}_{dc}=\frac{f}{T}0.19 \, \, \tilde{l}^{3.54}
\end{equation}
This scaling is considered to be valid only in a certain regime. On one side, the regime is limited by 
$\tilde{l} \geq 1.3$, since below that value substantial
parts of the spectrum become localized, which is, of course, expected to change transport behavior drastically. On the other side, this 
scaling is not necessarily believed to hold arbitrarily far from the critical point. Our numerics, however, indicate that it holds within
the regime displayed in the figures.
This is our first main quantitative result.
While the conductivities of the two model\_types appear to coincide at $\tilde{l}\approx 1.3$ the 
conductivity increases much faster with increasing mean overlap lengths in model\_type II. This supports
the concept of model\_type II exhibiting
Boltzmann transport for longer mean overlap lengths, while model\_type III always shows
non-Boltzmann transport.
\section{Diffusion constant and Einstein relation}
\label{seceinstein}
Apart from the conductivity the diffusion coefficient is another important transport
quantity. According to the Einstein relation, conductivity and the diffusion constant
should be proportional to each other. However, the validity of the Einstein relation
and the limits of its applicability have been  much debated subjects and continue to
be so in the context of quantum systems \cite{Steinigeweg2009} (and references
therein). Recently it has been reported that the Einstein relation holds for
periodic, interacting, 1D quantum systems at high temperatures. It is claimed to
hold even for finite times, thus taking the form \cite{Steinigeweg2009} 
\begin{equation}
\label{einsteinepl}
D(t)=\frac{T}{\epsilon^2} \sigma(t)
\end{equation}
where $D(t)$ is the (time-dependent) diffusion constant and $\epsilon^2$ is the
uncertainty (variance) of the transported quantity per site at the respective
equilibrium. In the following we investigate whether this relation also holds for
the disordered, non interacting, 3D quantum systems at hand. In our case the
transported quantity is the particle density. In the limit of high temperatures and
low fillings the equilibrium fluctuations scale as $\epsilon^2=f$ \cite{Kubo1991}.
Thus, if one hypothetically accepts the validity of (\ref{einsteinepl}) also for the
system at hand, one gets from inserting (\ref{kubo}) 
\begin{equation}
\label{einsteinone}
D(t)=\int_0^{t}\frac{1}{N}\text{Tr}\{ \hat{J}(t')\hat{J}(0)\} dt'
\end{equation}
In the following we demonstrate that almost the same relation between the diffusion
constant and the current auto correlation function may also be obtained from another
consideration which applies to the disordered systems at hand. If a diffusion
equation holds, the derivative with respect to time of the spatial variance of the diffusing
quantity equals twice the diffusion constant \cite{Steinigeweg2009}. We may analyze
the dynamics of this variance on quantum mechanical grounds for the models hat hand.
If initially the particle is completely concentrated at site $i$ and we assume that
the spatial expectation value does not move (which is an assumption since the model
is disordered, but since the disorder is isotropic the assumption appears reasonable
and  furthermore may be justified by numerical checking) the variance $\delta^2
x_i(t)$ reads
\begin{equation}
\label{vari}
\delta^2 x_i(t) = \sum_j(x_j-x_i)^2\text{Tr}\{\hat{n}_j(t)\hat{n}_i\} 
\end{equation}
Averaging this over all sites yields 
\begin{equation}
\label{var}
\Delta^2 x(t) = \frac{1}{N}\sum_{ij}(x_j-x_i)^2\text{Tr}\{\hat{n}_j(t)\hat{n}_i\} 
\end{equation}
Taking the second derivative with respect to time yields
\begin{equation}
\label{vara}
\frac{d^2}{dt^2}\Delta^2 x(t) = -\frac{1}{N}\sum_{ij}(x_j-x_i)^2\text{Tr}\{ 
[\hat{H} ,[\hat{H} ,\hat{n}_j(t)]]\hat{n}_i\} 
\end{equation}
Due to the invariance of the trace under cyclic permutation of the traced operators
this may be rewritten as 
\begin{equation}
\label{varb}
\frac{d^2}{dt^2}\Delta^2 x(t) = \frac{1}{N}\sum_{ij}(x_j-x_i)^2\text{Tr}\{ [\hat{H}
,\hat{n}_j(t)][\hat{H} ,\hat{n}_i]\} 
\end{equation}
Since the total particle number $\sum_i \hat{n}_i(t)$ is conserved, the respective
commutators vanish and the remainder reads
\begin{equation}
\label{varc}
\frac{d^2}{dt^2}\Delta^2 x(t) = \frac{-2}{N}\sum_{ij}\text{Tr}\{ [\hat{H}
,x_j\hat{n}_j(t)][\hat{H} ,x_i\hat{n}_i]\} 
\end{equation}
This, however, is essentially the velocity auto correlation function, cf.
(\ref{vel}) and (\ref{pos}), such that 
\begin{equation}
\label{vard}
\frac{d^2}{dt^2}\Delta^2 x(t) = \frac{2}{N}\text{Tr}\{ \hat{v}(t)\hat{v}(0)\} 
\end{equation}
Given that, as explained above, $\frac{d}{dt}\Delta^2 x(t) = 2D(t)$ we get
\begin{equation}
\label{einsteintwo}
D(t)=\int_0^{t}\frac{1}{N}\text{Tr}\{ \hat{v}(t')\hat{v}(0)\} dt'   
\end{equation}
Which is, up to the difference between  $\hat{v}$ and $\hat{J}$, the same relation
one also gets from boldly applying an Einstein relation that has been derived in a
different context, cf. (\ref{einsteinone}). The difference between  $\hat{v}$ and
$\hat{J}$ is not completely trivial since the periodic boundary conditions change
the topology of the system. Nevertheless, (\ref{einsteinone}) and (\ref{einsteintwo})
encourage a numerical check of the validity of the Einstein relation for the systems
at hand in the respective sense. This numerical check proceeds as follows: We
implement an initial state of the form
\begin{equation}
\label{init}
 \rho(0) =\frac{1}{Z}\exp{(-\frac{(\hat{x}-\frac{L}{2})^2}{2})}, \quad 
Z= \text{Tr}\{ \exp{(-\frac{(\hat{x}-\frac{L}{2})^2}{2})}    \}
\end{equation}
i.e., a state in which the probability is more or less concentrated in a thin slab
of a thickness on the order of 1, perpendicular to the $x$\_axis in the middle of
the cubic sample. We calculate the increase of the variance of this state and take a
derivative with respect to time,
\begin{equation}
\label{diffone}
D_1(t)=\frac{1}{2}\frac{d}{dt}\text{Tr}\{ \hat{x}^2(t) \rho(0) \}
\end{equation}
This corresponds to the diffusion constant one obtains from monitoring the spatial
expansion of the probability distribution. We compare this to the integrated current
auto correlation function, which is what (\ref{einsteinone}) and (\ref{einsteintwo})
imply:
\begin{equation}
\label{difftwo}
D_2(t)=\int_0^{t}\frac{1}{N}\text{Tr}\{ \hat{J}(t')\hat{J}(0)\} dt' 
\end{equation}
\begin{figure}[htbp]
\subfigure[]{\includegraphics[width=7.5cm]{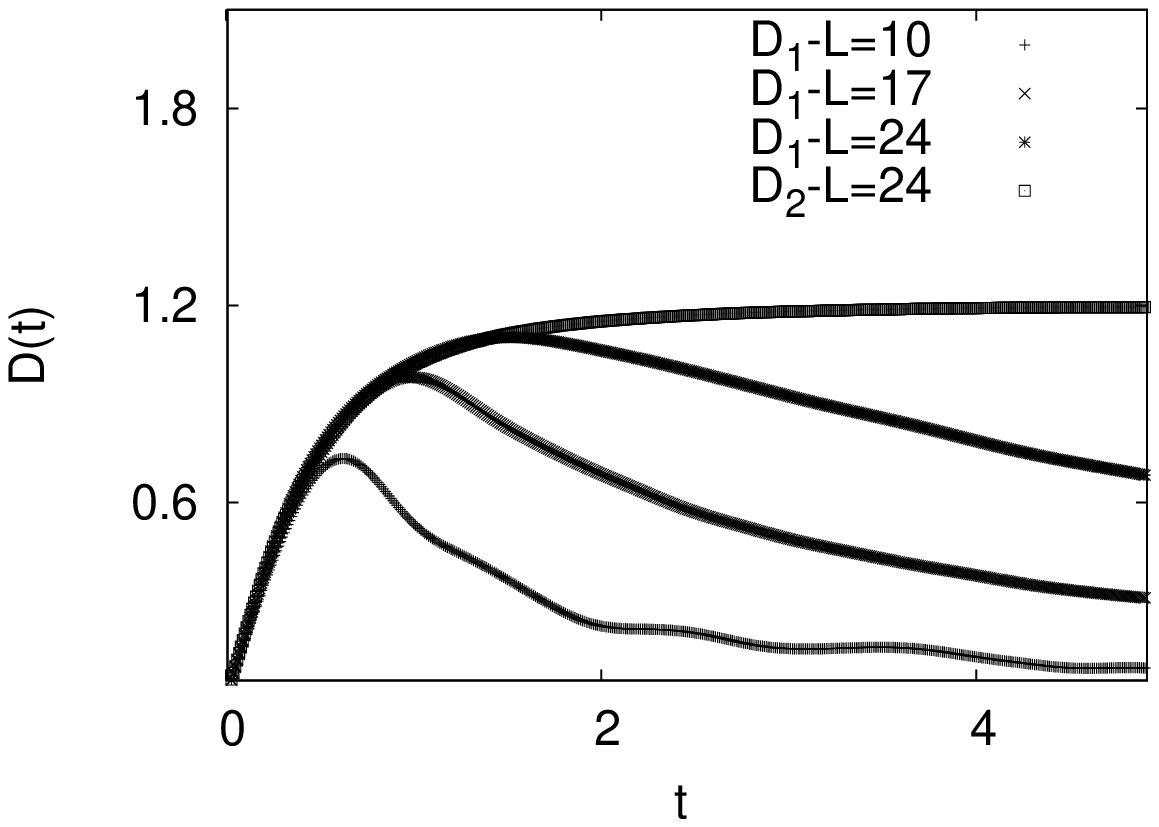}
\label{fig-9a}
}
\subfigure[]{\includegraphics[width=7.5cm]{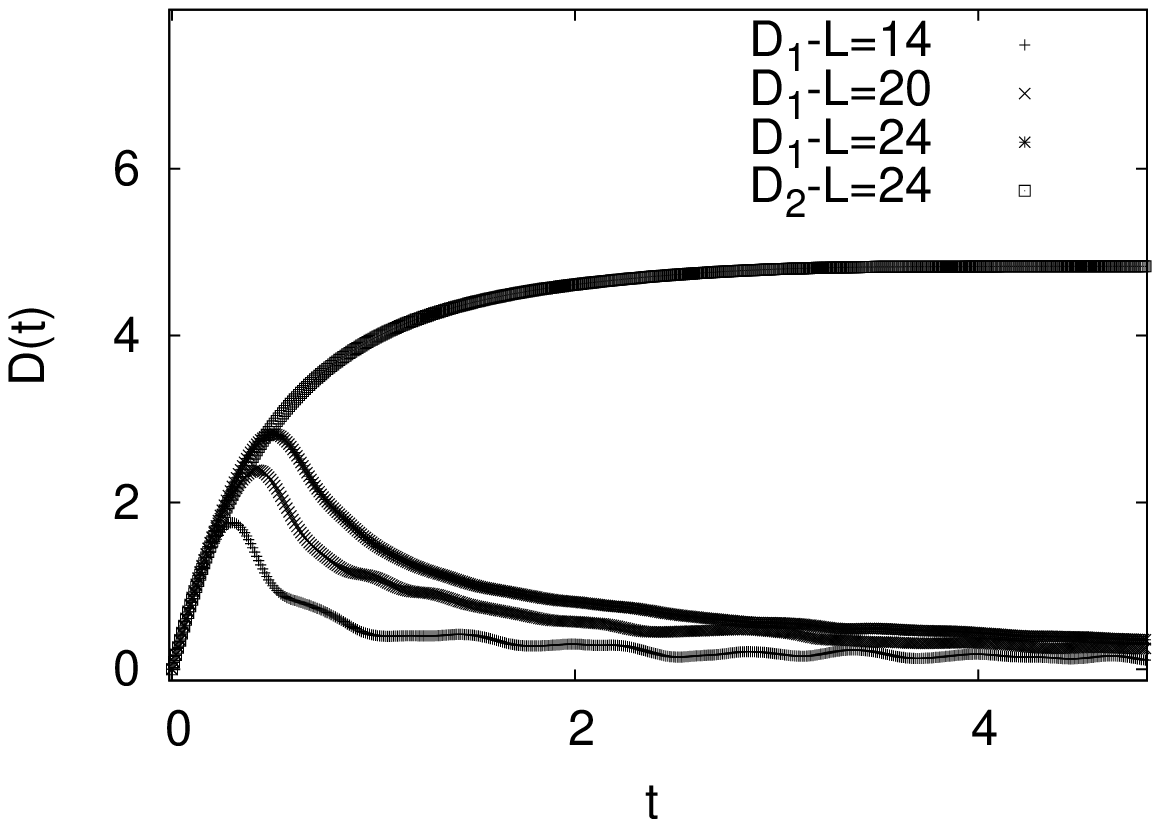}
\label{fig-9b}
}
\caption{Comparison of two methods to calculate (time-dependent) diffusion
coefficients: $D_1(t)$ from (\ref{diffone}) and   $D_2(t)$ from (\ref{difftwo}). The
data address model\_type II with (a) mean overlap length $\tilde{l}=1.5$ and (b)
mean overlap length $\tilde{l}=2.0$ for the indicated sample sizes. Obviously, finite-size
effects are more pronounced for $D_1(t)$; however, it appears to converge against
$D_2(t)$ for large sample sizes. This coincidence implies the validity of an
Einstein relation.}
\label{fig-9}
\end{figure}
\begin{figure}[htbp]
    \subfigure[]{\includegraphics[width=7.5cm]{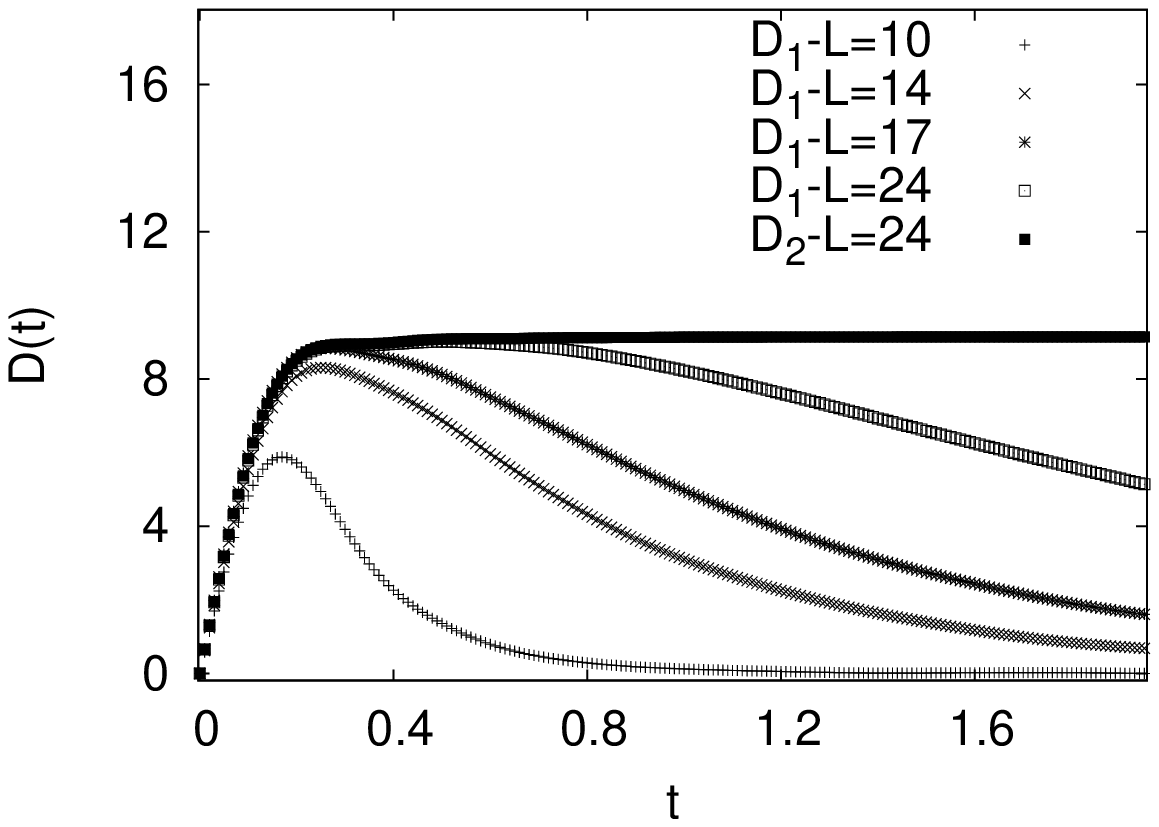}
\label{fig-10a}
}
    \subfigure[]{\includegraphics[width=7.5cm]{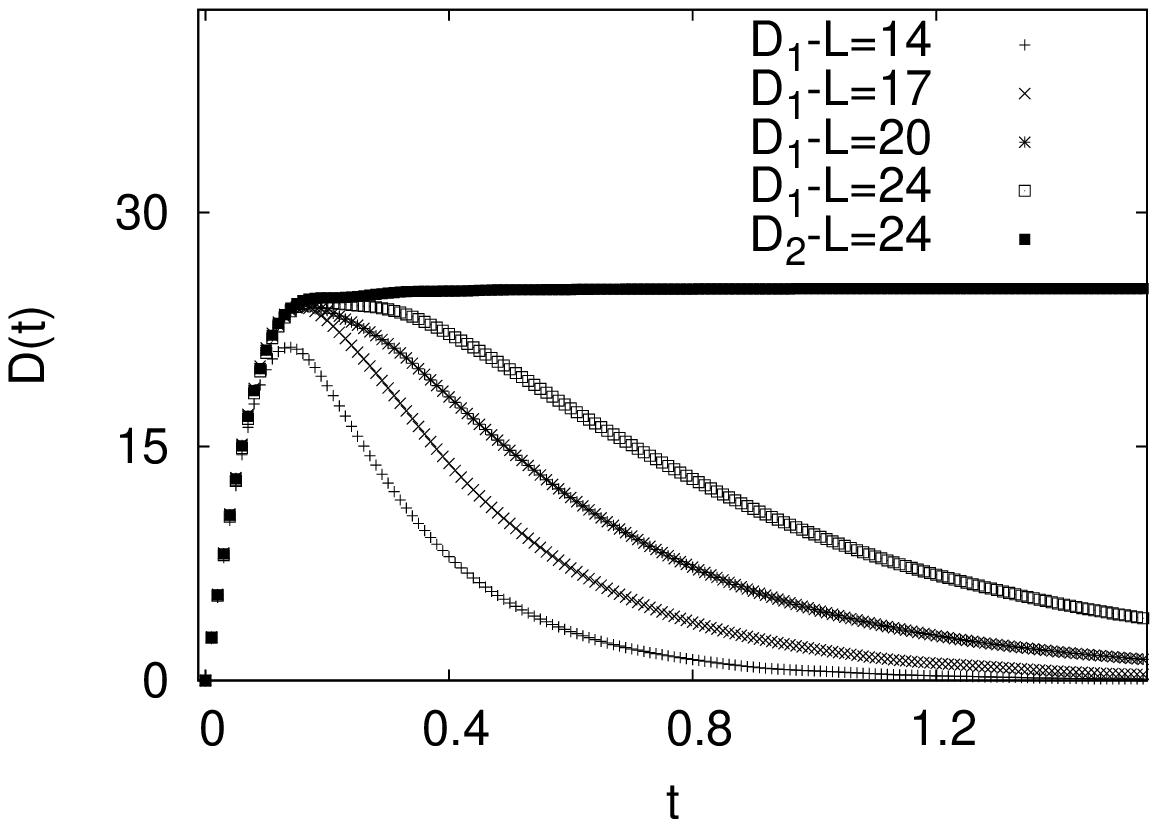}
\label{fig-10b}
}
\caption{Comparison of two methods to calculate (time-dependent) diffusion
coefficients: $D_1(t)$ from (\ref{diffone}) and   $D_2(t)$ from (\ref{difftwo}). The
data address model\_type III with (a) mean overlap length  $\tilde{l}=3.0$ and (b)
mean overlap length $\tilde{l}=4.0$ for the indicated sample sizes. Obviously, finite-size
effects are more pronounced for $D_1(t)$; however, it appears to converge against
$D_2(t)$ for large sample sizes. This coincidence implies the validity of an
Einstein relation. }
\label{fig-10}
\end{figure}
The results are displayed in Figs. \ref{fig-9} and \ref{fig-10} for model\_types II and
III. Although finite-size effects are much more pronounced if the diffusion constant
is calculated by means of (\ref{diffone}), there is, for the initial, valid time
period  a very good agreement in the sense of (\ref{einsteinepl}). Thus, we conclude
that the Einstein relation is valid for the systems at hand and Fig. \ref{fig-8} may
be viewed as not only describing the conductivity but also the diffusion constant at
high temperatures.
\section{Different types of transport behavior: Boltzmann- and NON-BOLTZMANN transport}
\label{transtype}
\begin{figure}[htbp]
\centering
\includegraphics[width=7.5cm]{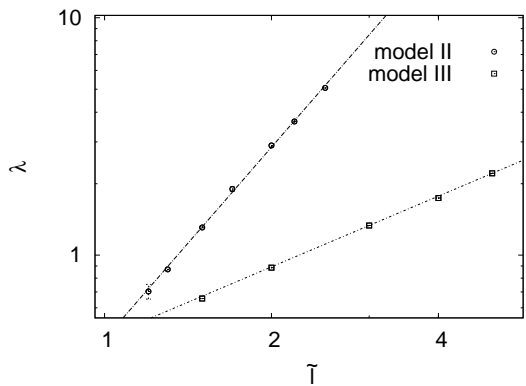}
\caption{Mean free paths $\lambda$ (for definition see text) for model\_types II and
III as a function of mean overlap length  $\tilde{l}$, sample size $L=26$. For both
model\_types the mean free paths appear to scale as  power laws with $\tilde{l}$;
the dashed ($---$, model\_type II) and dotted ($...$, model\_type III) lines are the
respective fits, cf. (\ref{free}). While for model\_type III the mean free path is
always lower than the mean overlap length $\lambda_{III}< \tilde{l}$ the mean free
path of model\_type II becomes larger than the mean overlap length $\tilde{l}$ at
about  $\tilde{l}\approx 1.8$. This indicates that model\_type II undergoes a
transition from non-Boltzmann to Boltzmann transport, while model\_type III does not. The
vertical error bars correspond to different random implementations of the same
model. These variations appear to increase for small $\tilde{l}$.} 
\label{fig-11}
\end{figure}

As already indicated in Sec. \ref{secconduct}, it appears reasonable to interpret the
described transport behavior in terms of two different transport types,  although both correspond to the metallic regime:
Non-Boltzmann transport which is (in a sense described below) comparable to the dynamics of an over damped
Brownian particle or the  thermally activated hopping transport which may occur in the localized regime of amorphous and/or doped semiconductors \cite{Shklovskii1984} and Boltzmann transport which resembles the dynamics of a particle
in a periodic lattice featuring some impurities or a system of quasi free, weakly interacting particles. These transport\_types have also
been found in other one-particle quantum systems, e.g., non-Boltzmann transport  in modular
quantum systems \cite{Weaver2006, Michel2005} and both transport-types in the 3D
Anderson model \cite{Steinigeweg2010, Brndiar2006}. In order to elaborate on this point somewhat
further we define a ``mean free path'' for the models at hand from the following
consideration: If the particle was completely ballistic (infinite mean free path)
the current auto correlation function would never decay and the time-dependent
diffusion coefficients in the sense of (\ref{einsteinone}) would always increase
linearly. The time-dependent diffusion 
coefficients 
of the models at hand increase linearly at the beginning, cf. Figs.
\ref{fig-9} and \ref{fig-10}, but reach a final plateau after that initial period. We
define, somewhat arbitrarily, the ballistic period as the period before the
diffusion coefficient has reached $90\%$ of its eventual value. Now we call the mean
free path the square root of the increase of the spatial variance of an initial
state of type (\ref{init}) during this ballistic period. So the mean free path is
roughly the initial increase of  width of an initially  narrow probability
distribution up to the point where the fully diffusive dynamics begins. The so-defined mean free paths $\lambda$ are displayed in Fig. \ref{fig-11}. The mean free
path appears to scale as a power law with the mean overlap length for both model\_types
II and III. The respective fits yield
\begin{equation}
\label{free}
\lambda_{II}=0.44 \, \, \tilde{l}^{2.68}, \quad  \lambda_{III}=0.45 \, \,
\tilde{l}^{0.99}
\end{equation}
 As already mentioned below (\ref{conduct}) this scaling is considered to be valid only in a certain regime.
On one side, the regime is limited by 
$\tilde{l} \geq 1.3$, since below that value substantial
parts of the spectrum become localized, which is of course expected to change transport behavior drastically. On the other side, this 
scaling is not necessarily believed to hold arbitrarily far from the critical point. Our numerics, however, indicate that it holds within
the regime displayed in the Figures.

This is our second main quantitative result. 
While the mean free paths  of the two model\_types are very similar for $\tilde{l} \approx 1.3$
[cf. Fig.
\ref{fig-5b}], the mean free path increases much faster with increasing hopping
lengths in model\_type II. 
This finding supports the classification of the two different types of transport: While
for model\_type III the mean free path $\lambda$ remains below and scales as the mean overlap
length for all  $\tilde{l}$, it appears that $\lambda$ becomes larger than
$\tilde{l}$ for mean overlap lengths above, say, $\tilde{l} \approx 1.8$ for model\_type
II. Thus, transport in model\_type III may always be classified as non-Boltzmann transport,
whereas the transport behavior of model\_type II appears to undergo a transition from
non-Boltzmann  to Boltzmann transport at about  $\tilde{l} \approx 1.8$. This point of
view is, furthermore, supported by the fact that the current auto correlation function
decays in Gaussian fashion at all $\tilde{l}$ for model\_type III, whereas it
undergoes a transition from Gaussian to exponential decay for model\_type II at
$\tilde{l} \approx 1.8$, cf. Fig. \ref{fig-7a}. Note that this Boltzmann transport
occurs, although model II is also topologically completely disordered, i.e.,
features no site order whatsoever.
\section{Summary and conclusion}
\label{sumcon}
We investigated the transport behavior of quantum systems which may be described as
three-dimensional, topologically completely disordered, one-particle, tight-binding
models on the basis of the Schroedinger equation. These models are intended as very
simplified descriptions of (hypothetical) solids in which the atomic nuclei are
distributed completely at random in space without any short- or long-range order. The
hopping or orbital overlap terms of the tight-binding Hamiltonian are simply taken to be decreasing
functions of the distances between the nuclei. By means of a simple method based on
the inverse participation number, we identify (rather roughly) the localized
regimes. While the Anderson transition appears to occur approximately at a mean
overlap length of $\tilde{l}\approx0.6$ (with respect to the mean site distance) for all
considered models, some models exhibit delocalized eigenstates at more or less all
energies already at mean overlap lengths of $\tilde{l}>1.3$ in others $30\%$ of the
spectrum remain localized up 
to mean overlap lengths of $\tilde{l} > 6$. For quantitative transport
investigations we focused on models which are almost entirely delocalized.  The
conductivity at low fillings and high temperatures has been determined by evaluating
the Kubo formula using numerically exact diagonalization for finite samples. It
turns out that valid quantitative results with negligible finite size effects may be
obtained for a range of models at sample sizes of about $17 000$ sites. The
conductivities  are  found to depend as power laws on the mean overlap lengths sufficiently above the Anderson transition. In
addition to the conductivity also the diffusion coefficient is addressed.
Theoretical considerations which suggest that an Einstein relation should hold,
i.e., that the diffusion coefficient may expected to be proportional to the
conductivity are presented. Those considerations are numerically confirmed by
monitoring the expansion of an initially narrow wave package. Eventually a mean free
path is defined and numerically determined. It is found that for a 
range of models the mean free path substantially exceeds the mean overlap length.
This suggests that these models  may be thought of as systems in which particles
travel almost ballistically over distances much larger than the typical site
distance and undergo only weak scattering, like in the case of a periodic crystal
containing some impurities. This holds although the systems feature no spatial order
whatsoever.
\newpage

%

\end{document}